\documentclass{IEEEtran}
\usepackage{cite}
\usepackage{amsmath,amssymb,amsfonts}
\usepackage{algorithmic}
\usepackage{graphicx}
\usepackage{textcomp}
\usepackage{wrapfig}
\usepackage[utf8]{inputenc}
\usepackage{amsfonts,amssymb,mathtools}
\usepackage{array}
\usepackage{graphicx}
\usepackage{multirow}
\usepackage{multicol}
\usepackage{soul}
\usepackage{balance} 
\usepackage{booktabs}
\usepackage{upgreek}

\begin{document}
\title{Probabilistic Allocation of Payload Code Rate and Header Copies in LR-FHSS Networks}
\author{{Jamil de Araujo Farhat, Jean Michel de Souza Sant'Ana, João~Luiz~Rebelatto,  Nurul~Huda~Mahmood, Gianni~Pasolini and Richard~Demo~Souza}
\thanks{J. Farhat and J. L. Rebelatto are with the Federal University of Technology-Paraná (UTFPR), Curitiba, Brazil, {\{jamilfarhat, jlrebelatto\}@utfpr.edu.br}. 
J. M. S. Sant'Ana and N. H. Mahmood are with the Centre of Wireless Communications (CWC), University of Oulu, Oulu, Finland, {\{jean.desouzasantana, nurulhuda.mahmood\}@oulu.fi}. 
G. Pasolini is with the Department of Electrical, Electronic, and Information Engineering, University of Bologna, Bologna, Italy, {gianni.pasolini@unibo.it}.
R. D. Souza is with the Department of Electrical and Electronics Engineering, Federal University of Santa Catarina, Florianópolis, Brazil, { richard.demo@ufsc.br}. }}

\maketitle

\begin{abstract}
We evaluate the performance of the LoRaWAN Long-Range Frequency Hopping Spread Spectrum (LR-FHSS) technique using a device-level probabilistic strategy for code rate and header replica allocation. Specifically, we investigate the effects of different header replica and code rate allocations at each end-device, guided by a probability distribution provided by the network server. As a benchmark, we compare the proposed strategy with the standardized LR-FHSS data rates DR8 and DR9. Our numerical results demonstrate that the proposed strategy consistently outperforms the DR8 and DR9 standard data rates across all considered scenarios. Notably, our findings reveal that the optimal distribution rarely includes data rate DR9, while data rate DR8 significantly contributes to the goodput and energy efficiency optimizations.
\end{abstract}


\maketitle

\section{Introduction}
\label{sec:introduction}

\IEEEPARstart{T}{he} rapid growth of Internet of Things (IoT) spurred applications across diverse domains, including agriculture, industry, environmental monitoring, and smart cities~\cite{Arun.2024, Guo.2024}, each with its own unique set of requirements. Within this diverse ecosystem, one particularly challenging scenario is related to massive IoT (mIoT). This involves applications that support a vast number of network devices with constraints of low power consumption and complexity, while tolerating reduced data rates~\cite{6g_flagship, Jouhari.2023}. In response to these demands, Low-Power Wide Area Networks (LPWANs) emerge as a collection of technologies designed to meet mIoT requirements. LPWANs enable long-distance communication that is also energy-efficient, aligning with the goals of mIoT~\cite{Pandey.2022, Pérez.2022, Fraire.2022, Azim.2024}.

Among various alternatives for addressing such challenging scenarios, LoRa emerges as a remarkable solution~\cite{Shanmuga.2020, Jiang.2022}. This technology, patented by Semtech~\cite{Semtech}, utilizes Chirp Spread Spectrum (CSS) and achieves a balance between data rate and receiver sensitivity through the use of different spreading factors~\cite{Pasolini_21}. Moreover, LoRa typically operates within a star topology under the LoRaWAN protocol~\cite{Raza.2017, Chilamkurthy.2022}. In this architecture, the end-devices communicate directly with a gateway using an Aloha-based protocol, which operates without any collision prevention mechanism.

A recent expansion of the LoRaWAN specification~\cite{LoRaWanParameters} has incorporated the Long Range Frequency Hopping Spread Spectrum (LR-FHSS) technique in the uplink to support an increased number of end-devices~\cite{Boquet.2021, Herrería.2024}, with improved robustness to interference~\cite{Lin.2023}. LR-FHSS, which falls into the category of Telegram Splitting Multiple Access (TSMA) techniques~\cite{ETSI_TS_103_357}, divides the coded payload into small segments known as fragments, which are then transmitted in a pseudo-random fashion across various frequency hops. Additionally, prior to the transmission of payload fragments, multiple replicas of the header are sent using different frequencies~\cite{Alvarez.2022}. This strategy exploits diversity to improve the packet transmission success rate, while maintaining similar, if not larger, communication range as in standard LoRa~\cite{Cheikh.2022}. 

The LR-FHSS modulation has recently gained significant attention due to its relevance in the scope of mIoT~\cite{Boquet.2021, Maleki.IEEE_CL.2024, Ullah.2022, Alvarez.2022, Jung.2023, Wang.2024, SantAnna.2024, Knop.2024}.  An initial study of the LR-FHSS performance is reported in~\cite{Boquet.2021}, demonstrating that this technique outperforms conventional LoRa CSS modulation in massive network scenarios. Furthermore, an analytical approach to derive the outage probability of LR-FHSS in Direct-to-Satellite (DtS) networks is presented in~\cite{Maleki.IEEE_CL.2024}, considering diverse effects related to the wireless medium. The authors highlight a significant performance improvement in terms of network capacity compared to the conventional LoRa. Additionally, the packet delivery rate using analytical and simulation models is investigated in~\cite{Ullah.2022}, emphasizing the impact of the header success probability on the overall reliability of LR-FHSS.

Several approaches have been explored to enhance the performance of LR-FHSS networks. In the context of satellite networks, the trajectory information is leveraged in~\cite{Alvarez.2022} to propose uplink transmission policies that improve network scalability. Meanwhile, the authors in~\cite{Jung.2023} focus on developing an accurate framing scheme for LR-FHSS, in which a robust signal detector is proposed for the receiver. Additionally, the work in~\cite{Wang.2024} delves into different coded frequency hopping designs, incorporating segment-level coding with erasure detection. This enhancement improves the resilience against clustered errors that can be commonly encountered in LR-FHSS, ultimately enhancing communication reliability. 

Given that the LR-FHSS protocol is built upon Aloha with message replication~\cite{Susilo.2023}, strategies to improve the performance of any Aloha-based network could also prove beneficial within the context of LR-FHSS. Recently,~\cite{SantAnna.2024} has considered the asynchronous contention resolution diversity Aloha protocol to mitigate collisions. The proposed scheme demonstrates significant improvements in various network performance metrics. Furthermore, considering the dominant role of headers in the overall system performance, a network-coded-aided inter-header combination scheme is proposed in~\cite{Knop.2024}. The method aims at increasing header protection against both erasures and collisions, resulting in important improvements in terms of packet error rate under several scenarios. 

Most of the literature, as well as the LR-FHSS specification, exploit message replication for improved performance. Against this background, and considering the benefits of irregular repetition strategies in Aloha networks~\cite{Tralli.2024}, which exploit the diversity of repetition strategies among end-devices, we propose a device-level probabilistic strategy for payload code rate and header replica allocation. In this approach, we define a varied set of replicas and code rate configurations for end-devices, aiming to improve key network performance indicators such as goodput and energy efficiency in IoT applications with numerous users. The end-devices randomly select configuration setups for each transmission based on a probability distribution provided by the network server. An important aspect of this work is that we consider only the code rates available in current LR-FHSS-enabled chipsets, ensuring compatibility with the existing infrastructure. Moreover, it is worth mentioning that the proposed method can be incorporated into any of the previously mentioned related works.

The contributions of this work can be summarized as:
\begin{itemize}
    \item We analyze the impact of different probabilistic configurations (in terms of payload code rate and header replicas) in the network goodput and energy efficiency, for different numbers of end-devices and payload sizes.
    \item Through an extensive analysis, we show that the proposed probabilistic approach outperforms the standardized data rates, both in terms of goodput and energy efficiency.
    \item The proposed approach achieves near-optimal results with low quantization resolution of the probability distribution, demonstrating that only three bits of feedback at the downlink are sufficient to inform the end-devices. 
\end{itemize}

The remainder of this document is organized as follows. Section~\ref{sec:Preliminaries} briefly reviews the LR-FHSS specification. Section~\ref{sec:DistributedStrategy} discusses the proposed strategy, while Section~\ref{sec:NumericalResults} presents some numerical results to evaluate its performance. Finally, Section~\ref{sec:Conclusions} concludes the paper.

\section{Preliminaries}
\label{sec:Preliminaries}

LR-FHSS is a modulation technique recently added to the LoRaWAN specification~\cite{Wang.2024}. Its primary purpose is to enhance uplink capacity, allowing improved performance in dense networks served by a single gateway, such as the case of DtS IoT networks~\cite{Semtech.AppNote}. In LR-FHSS, the total available bandwidth is segmented into many multiple physical channels, each with a bandwidth of 488 Hz~\cite{SantAnna.2024}. Furthermore, these physical channels are organized into time-frequency grids, with $c$ physical channels per grid.

\begin{figure}[!t] 
\centering
\includegraphics[width=8.8cm]{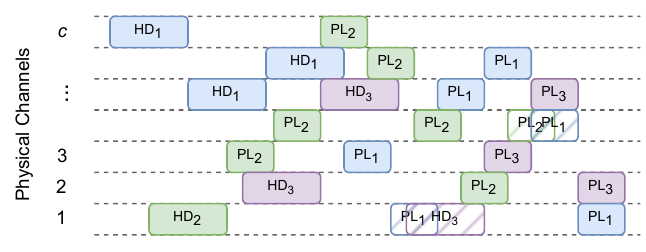}
\caption{End-devices employ varied configurations of header replicas and coding rates to transmit LR-FHSS packets across $c$ physical channels in the frequency domain. The subscript on each header replica ($\mathrm{HD}_m$) and payload fragment ($\mathrm{PL}_m$) indicates its association with the $m$-th user, while the striped boxes illustrate collisions between different devices.}
\label{Fig:LRFHSS_packet}
\end{figure}

The transmission of a LR-FHSS packet involves both a header and a payload transmission. To boost robustness against collision events, an end-device sends multiple header copies sequentially across different physical channels, as illustrated in Fig.~\ref{Fig:LRFHSS_packet}. Additionally, the payload is coded by a convolutional code of a certain rate. The coded payload is then divided into several fragments, which are transmitted sequentially over different physical channels within the grid. 
Each header replica has a Time-on-Air (ToA) of $t_h=233.472$~ms, representing the time for each frequency hop. Furthermore, the ToA for each payload fragment is $t_{f}=102.4$~ms~\cite{LoRaWanParameters}. Let us remark that the LR-FHSS technique includes various data rate (DR) models, each associated with a specific pair of payload code rate and number of header replicas.

\begin{figure}[!t] 
\centering
\includegraphics[width=8.8cm]{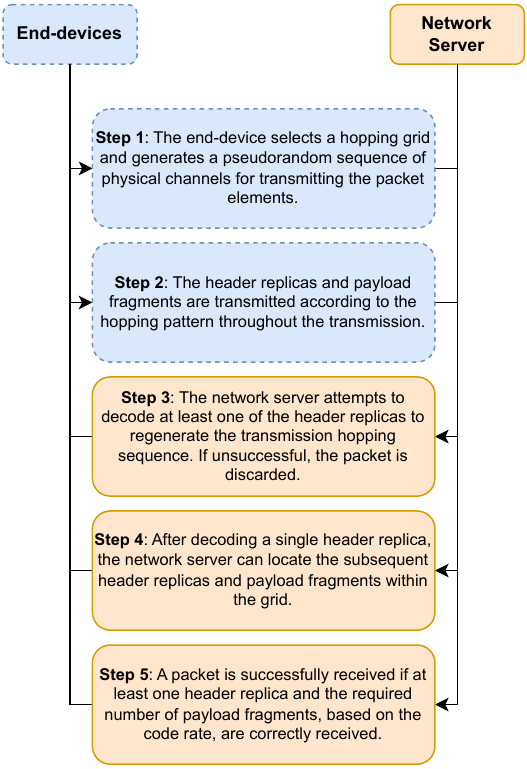}
\caption{Overview of the steps involved in LR-FHSS packet communication process between an end-device and the network server.}
\label{Fig:Flowchart_LRFHSS}
\end{figure}

An overview of the LR-FHSS packet communication process is provided in Fig.~\ref{Fig:Flowchart_LRFHSS}. The process begins with an end-device selecting a hopping grid and then randomly picking a seed to generate a pseudorandom sequence of physical channels prior to the transmission of the first header replica, as shown in \textbf{Step 1}. During transmission, the end-device hops to a new physical channel for every header replica and payload fragment, changing the carrier frequency according to the hopping sequence, as illustrated in \textbf{Step 2}. The header includes crucial information, such as an indicator for generating the pseudorandom frequency hopping sequence. To regenerate this sequence, the gateway must decode at least one header replica, as described in \textbf{Step 3}, enabling it to locate the payload fragments within the grid, as depicted in \textbf{Step 4}. Additionally, a minimum number of payload fragments, determined by the code rate in use, must be decoded for successful message recovery, as illustrated in \textbf{Step 5}.

\section{Proposed Strategy}
\label{sec:DistributedStrategy}

We consider an LR-FHSS network with $M$ end-devices, using transmit power $\mathcal{P}_t$ to send $l$ bytes of payload to a single gateway, which is connected to a network server. Furthermore, the time between  transmissions from each end-device follows an exponential random variable $z$ with mean $1/\lambda$. This implies that, on average, $\lambda$ transmissions occur per second, with inter-arrival time probability distribution $\lambda\,\mathrm{e}^{-\lambda\,z}$, for $z>0$.

\begin{figure}[!t] 
\centering
\includegraphics[width=7.5cm]{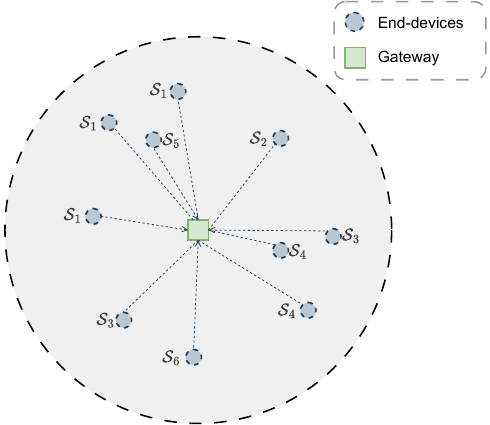}
\caption{Network with $M\!=\!10$ end-devices communicating to a gateway, with each end-device using a transmission setup $\mathcal{S}_k$, $k \in \{1,\ldots,6\}$.}
\label{Fig:SystemModel}
\end{figure}

According to the strategy we propose, for each transmission attempt, the end-devices randomly choose a setup $\mathcal{S}_k$, with ${k \in \{1, \dots, K\}}$, associated with a number of header replicas and payload code rates. Moreover, the network server selects a set $\Updelta=\left\{\delta_1, \dots, \delta_K\right\}$, with $\sum_{k=1}^{K} \delta_k=1$, which corresponds to a probability distribution that should be respected by the end-devices when choosing a transmission setup. In other words, $\delta_k$ is the assignment probability for each $\mathcal{S}_k$\footnote{The transmission of this probability distribution to the end-devices could be made by means of a multicast strategy or by simply piggybacking it in the downlink messages for Class A LoRaWAN devices. It is important to highlight that all LoRaWAN devices must support Class A implementation, which involves two short receive windows for downlink messages from the network once the uplink transmission is completed~\cite{LoRaWanSpecification}.}. A high-level illustration of the proposed strategy is shown in Fig.~\ref{Fig:SystemModel}, for $M=10$ end-devices and $K=6$ possible setups of header replicas and payload code rates.

Following a methodology similar to~\cite{Ullah.2022}, we write the average number of header and payload fragment transmissions in the network per unit of time respectively as
\begin{equation}
    {\lambda}_h = \bar{h}\,\lambda\,M
\end{equation} 
and
\begin{equation}
    {\lambda}_f = \bar{f}\,\lambda\,M,
\end{equation}where $\bar{h}$ represents the average number of header replicas per message, taking into account the proportion of end-devices using each specific setup, which is expressed as
\begin{equation}
    \bar{h} = \sum_{k=1}^{K} h_k \, \delta_k,
\end{equation}
where $h_k$ is the number of header replicas associated with $\mathcal{S}_k$. Moreover, $\bar{f}$ is the average number of coded payload fragments, which can be obtained as
\begin{equation}
    \bar{f} = \sum_{k=1}^{K} f_k \, \delta_k,
\end{equation}
where $f_k$ depends on the code rate and can be obtained as~\cite{LoRaWanParameters}
\begin{equation}
    f_k=\left[\frac{l+3}{6\,\mathrm{CR}_k}\right]
    \label{eq:payload_fragments}
\end{equation}
where $\mathrm{CR}_k$ is the code rate. Then, the average ToA of a message across the network is $\bar{h}\,t_h+\bar{f}\,t_f$.

We assume that the devices are within the coverage of the gateway, so that decoding errors are strongly dominated by potential collisions with transmission from different devices. Such assumption is reasonable in mIoT scenarios limited by interference, and it has been often considered in the literature~\cite{Paul.2020, Boquet.2021, Xiao.2022, Ullah.2022, santana2024lrfhsssim, SantAnna.2024}, as a good compromise between mathematical tractability and precision. Then, for a header replica transmitted at time $x_m$ for a specific end-device, we define its vulnerability interval as ${(x_m - t_h, x_m + t_h)}$. Assuming the same size for all payload fragments, we calculate the average number of packet elements, headers, and payload fragments that arrive during the vulnerable interval of a header replica, including itself, as 
\begin{equation}
    A_h=\max \left(1, {\lambda}_h\,2\,t_h + {\lambda}_f\left(t_h+t_f\right)\right),
    \label{eq:vulnerable_interval_header}
\end{equation}
where ${\lambda}_h\,2\,t_h$ is the average number of headers transmitted within the header vulnerable interval, while ${\lambda}_f\,\left(t_h+t_f\right)$ is the average number of payload fragments transmitted in the same interval. Additionally, the $\max(1,\cdot)$ operation is included to guarantee that $A_h\geq 1$, since we are assuming that at least one header replica has been transmitted. 

Similarly, for payload fragments, whose vulnerable interval is $(x_m - t_f, x_m + t_f)$, the average number of packet elements arriving at the gateway in the payload fragment vulnerable interval, including itself, is
\begin{equation}
    A_f=\max \left(1, {\lambda}_f\,2\,t_f + {\lambda}_h\left(t_h+t_f\right)\right).
    \label{eq:vulnerable_interval_fragment}
\end{equation}

Thus, an LR-FHSS packet is successfully recovered when the gateway correctly receives at least one of the $h_k$ headers, as well as an amount of $\mu_k = \lceil f_k \times \mathrm{CR}_k \rceil$ from the total $f_k$ payload fragments,  where $\lceil \cdot \rceil$ is the ceiling operator. 

Let us then refer to $P_{h, k}$ as the probability of successfully receiving at least one of the $h_k$ header replicas when employing  $\mathcal{S}_k$. Therefore, given that there are $c$ physical channels within the grid, the aforementioned probability is~\cite{Ullah.2022} 
\begin{equation}
    P_{h, k} = 1-\left(1-\left(1-\frac{1}{c}\right)^{A_h-1}\right)^{h_k}
    \label{eq:header_probability}
\end{equation}
where $\left(1-\frac{1}{c}\right)^{A_h-1}$ is the probability of correctly receiving a single header replica without collisions while considering the network average header vulnerable interval defined in~\eqref{eq:vulnerable_interval_header}. 

Similarly, the probability of successfully receiving a single payload fragment is 
\begin{equation}
    P_{f} = \left(1-\frac{1}{c}\right)^{A_f-1}.
\end{equation}
Therefore, the probability of successfully receiving at least $\mu_k$ payload fragments when adopting $\mathcal{S}_k$ is
\begin{equation}
    P_{\mu, k} = 1-\sum_{i=0}^{\mu_k-1} {f_k \choose i} {P_f}^i\left(1-P_f\right)^{f_k-i}
    \label{eq:CR_probability}
\end{equation}where ${a \choose b} = \frac{a!}{b!(a-b)!}$ is the binomial coefficient.

Finally, the overall packet success probability $P_s$ is determined by the weighted contribution of each setup $\mathcal{S}_k$ as
\begin{equation}
    P_s = \sum_{k=1}^{K} \delta_k\,\,P_{h, k}\,\,P_{\mu, k}.
    \label{eq:packet_success_probability}
\end{equation}

\subsection{Network Metrics}

The performance of the proposed strategy is evaluated in terms of two metrics, namely goodput and energy efficiency. The goodput, $\mathcal{G}$, represents the average number of successfully received bytes per second and is defined by
\begin{equation}
    \mathcal{G} = P_s\,M\,\lambda\,l \qquad \text{[bytes/s].}
    \label{eq:goodput}
\end{equation}
Another metric of interest is the energy efficiency, defined as
\begin{equation}
    \mathcal{E} = \frac{\mathcal{G}}{W} \qquad \text{[bytes/Joules],}
    \label{eq:energy_efficiency}
\end{equation}
which is the ratio of the goodput $\mathcal{G}$ to the average power $W$ expended by the end-devices, which, in turn, is 
\begin{equation}
    W = \mathcal{P}_t\,M\,\lambda\,(\bar{h}\,t_h+\bar{f}\,t_f) \qquad \text{[Joules/s]},
\end{equation}where $\lambda\,(\bar{h}\,t_h+\bar{f}\,t_f)$ corresponds to the fraction of time that a device is actively transmitting. 

\subsection{Optimization of $\Updelta$}

The optimization of a given metric $\mathcal{M} \in \{\mathcal{G}, \mathcal{E}\}$ involves finding the optimal distribution  $\Updelta^\star$, that maximizes the desired metric, which can be formally defined as 
\begin{subequations}
\begin{align}
\label{eq:OptimizationEquation}
\Updelta^\star &= \max_{\Updelta} \quad \mathcal{M} \\
\label{eq:constraintDeltaEquation}
&\text{s.t.} \, \sum_{k=1}^{K} {\delta_k}^\star=1,
\end{align}
\end{subequations}
where ${\delta_k}^\star$  is the optimal probability distribution related to $k$-th setup. However, due to the complexity of~\eqref{eq:goodput} and~\eqref{eq:energy_efficiency}, which depend on the configurations of header replicas and code rates, it is hard (if possible) to derive a closed-form equation for $\Updelta^\star$ that maximizes either the goodput or the energy efficiency. Notably, the objective functions exhibit neither concave nor convex behavior depending on the number of end-devices.

Thus, we perform such a task by means of an exhaustive search. This search is carried out on a discrete set of $\delta_k$ values, which the network server evaluates and communicates back to the end-devices. In terms of complexity, considering a scenario with $K$ different setups and $L$ possible values of $\delta_k$, the exhaustive search requires $L^K$ iterations, resulting in computational complexity $\mathcal{O}(L^K)$. However, it is important to note that this computational burden is confined to the network server and may vary depending on its hardware capabilities, while remaining transparent from the end-devices perspective.

\section{Numerical Results}
\label{sec:NumericalResults}

This section presents some numerical results to assess the performance of the proposed scheme. The evaluation was conducted using a discrete-event simulator implemented within the SciPy framework in Python, as detailed in~\cite{santana2024lrfhsssim}. Unless specified otherwise, we adopt $\mathcal{P}_t\!=\!20$~dBm, ${\lambda\!=\!1/900}$~s with a simulation duration of $3600$~s, $c\!=\!35$ channels and $l\!=\!10$~bytes. 

We compare the proposed strategy with approaches that employ LoRaWAN configurations DR8 or DR9, which are standard data rates specified in \cite{LoRaWanParameters}, as summarized in Table~\ref{table:DR_specification}. Moreover, based on the features of current LR-FHSS radios \cite{LR1121}, the proposed approach considers combinations of header replicas and payload code rates as listed in Table~\ref{table:set_combinations}. It is important to note that Table~\ref{table:set_combinations} defines a varied set of configuration setups to be employed by the end-devices, according to a probability distribution associated with each setup, guided by the network server. Moreover, note that according to the LoRaWAN parameters~\cite{LoRaWanParameters}, the standard DR8 model is equivalent to $\mathcal{S}_6$, while the DR9 model corresponds to $\mathcal{S}_3$, both highlighted in boldface in Table~\ref{table:set_combinations}. Therefore, our proposed strategy incorporates standard DR models to enhance performance by leveraging the benefits of diversity in the context of Aloha networks.

 \begin{table}[!t]
\renewcommand{\arraystretch}{1.5}
\caption{LoRaWAN standardized DR8 and DR9.}
\centering
\begin{tabular}{l|cc}
\toprule
\multicolumn{1}{c|}{LoRaWAN} & DR8 & DR9 \\ \midrule
Header Replicas ($\mathrm{h}$) & $3$ & $2$\\
Code Rate ($\mathrm{CR}$)  & $1/3$ & $2/3$ \\ \bottomrule
\end{tabular}
\label{table:DR_specification}
\end{table}

\begin{table}[!t]
\renewcommand{\arraystretch}{1.5}
\caption{Setups of Header Replicas and Code Rates.}
\centering
\begin{tabular}{l|cccccc}
\toprule
\multicolumn{1}{c|}{$\mathcal{S}_k$} & $\mathcal{S}_1$ & $\mathcal{S}_2$ & $\mathcal{S}_3$ & $\mathcal{S}_4$ & $\mathcal{S}_5$ & $\mathcal{S}_6$ \\ \midrule
Header Replicas ($h_k$) & $1$ & $1$ & $\mathbf{2}$ & $2$ & $3$ & $\mathbf{3}$ \\ 
Code Rate ($\mathrm{CR}_k$)  & $5/6$ & $2/3$ & $\mathbf{2/3}$ & $1/2$ & $1/2$ & $\mathbf{1/3}$ \\ \bottomrule
\end{tabular}
\label{table:set_combinations}
\end{table}

\subsection{Goodput Optimization}

\begin{figure}[!t] 
\centering
\includegraphics[width=8.8cm]{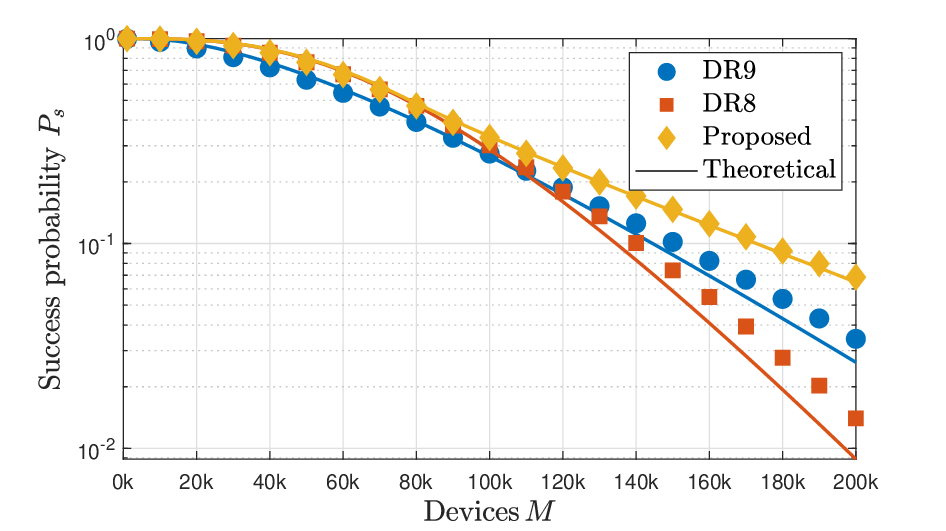}
\caption{Success probability versus the number of devices, for DR8, DR9, and the proposed method optimized for the maximum goodput.}
\label{fig:Probability_devices}
\end{figure}

\begin{figure}[!t] 
\centering
\includegraphics[width=8.8cm]{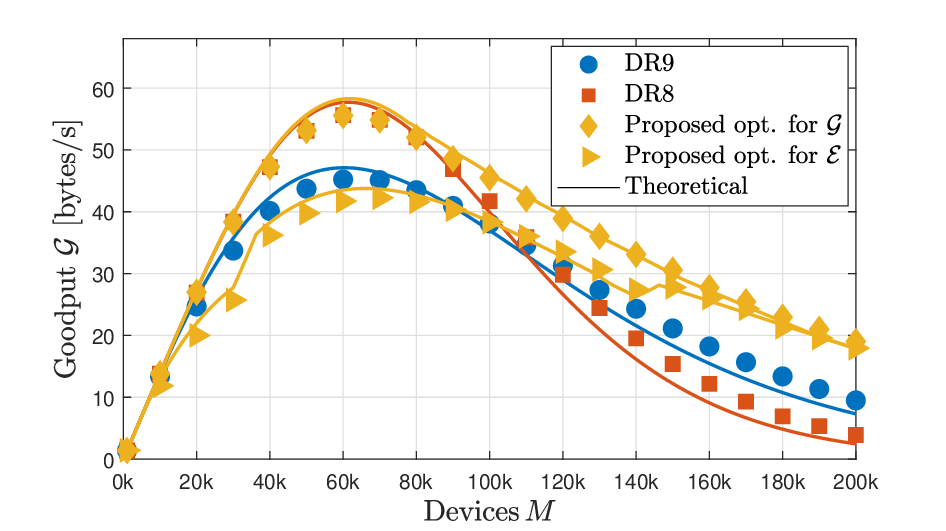}
\caption{Goodput versus the number of devices, for DR8, DR9, and the proposed method optimized for either maximum goodput or maximum energy efficiency.}
\label{fig:Goodput_devices}
\end{figure}

Initially, Fig.~\ref{fig:Probability_devices} illustrates the packet success probability, $P_{s}$, as a function of the number of devices, $M$. As we can observe, $P_{s}$ decreases with $M$, a consequence of the more congested scenario leading to increased collisions among users. Compared with standard DRs, the proposed method shows improved performance. Moreover, the theoretical expressions obtained in~\eqref{eq:packet_success_probability} are employed in the optimization problem given by~\eqref{eq:OptimizationEquation}. An exhaustive search is used to find the optimal $\Updelta$ that maximizes the goodput, and the results are in good agreement with the simulations\footnote{The discrepancies between theoretical and simulated results occur because the former ignores the correlation between header copies and payload fragments. When one fragment collides, the likelihood of another fragment from the same message also colliding increases. This interdependence is not considered in \eqref{eq:packet_success_probability}, where all transmissions are assumed to be independent.}. The proposed method considers the distribution $\Updelta$ that maximizes $P_s$ for the given number of devices, as will be detailed later. Let us note that, following~\eqref{eq:goodput}, for a given payload $l$ the optimization for $P_s$ is also an optimization for the goodput ${\cal G}$. Thus, Fig.~\ref{fig:Goodput_devices} plots the proposed strategy optimized for $\mathcal{G}$ versus $M$, showing its superior performance compared to standard DRs, especially in crowded scenarios. Additionally, the goodput of the strategy optimized for $\mathcal{E}$ is also presented, demonstrating that optimizing for energy efficiency does not guarantee maximum goodput.

To complement the analysis, Table~\ref{table:OptimalDistributionGoodput} presents the optimal distribution $\Updelta^{\star}$ of header replicas and code rate combinations to maximize $\mathcal{G}$ for the proposed strategy. One can see that for a relatively small number of end-devices, prioritizing a higher number of header replicas increases the packet success probability at the gateway. Thus, $\mathcal{S}_6$ (equivalent to DR8, with $h_6=3$ and $\mathrm{CR}_6= 1/3$) dominates in the $M<80000$ region, being chosen with probability $\delta_6=1$. In contrast, with an increase in the number of end-devices, it becomes crucial to find a balanced mix of header replicas and code rates. This is demonstrated by the balance between the $\mathcal{S}_1$ (with $h_1= 1$ and $\mathrm{CR}_1= 5/6$) and $\mathcal{S}_6$ setups. Notably, setups $\mathcal{S}_2$, $\mathcal{S}_3$ (equivalent to DR9), $\mathcal{S}_4$ and $\mathcal{S}_5$ have not been used in this scenario.

\begin{table}[!t]
\renewcommand{\arraystretch}{1.5}
\caption{$\Updelta^{\star}$ to maximize $\mathcal{G}$, for $l=10$ bytes [\%].}
\centering
\begin{tabular}{c|cccccc}
\toprule
$\mathbf{M}$ & $\delta_{1}$ & $\delta_{2}$ & $\delta_{3}$ & $\delta_{4}$ & $\delta_{5}$ & $\delta_{6}$ \\  
\midrule
$20000$& - & - & - & - & - & $100$\%  \\
$40000$ & - & - & - & - & - & $100$\%  \\
$60000$ & - & - & - & - & - & $100$\%  \\
$80000$ & $10$\% & - & - & - & - & $90$\%  \\
$100000$ & $35$\% & - & - & - & - & $65$\%  \\
$120000$ & $50$\% & - & - & - & - & $50$\%  \\
$140000$ & $60$\% & - & - & - & -& $40$\%  \\
$160000$ & $65$\% & - & - & - & - & $35$\%  \\
$180000$ & $75$\% & - & - & - & - & $25$\%  \\
$200000$ & $75$\% & - & - & - & - & $25$\% \\
\bottomrule
\end{tabular}
\label{table:OptimalDistributionGoodput}
\end{table}

\subsection{Energy Efficiency Optimization}

\begin{figure}[!t] 
\centering
\includegraphics[width=8.8cm]{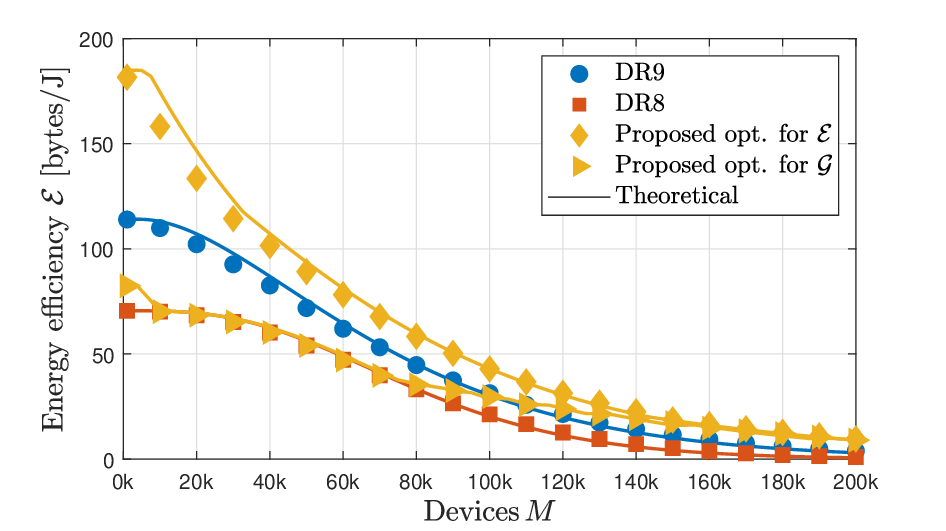}
\caption{Energy efficiency versus the number of devices, for DR8, DR9, and the proposed method optimized for either maximum energy efficiency or maximum goodput.}
\label{fig:MaximumEfficiency}
\end{figure}

In this section, we focus on maximizing the energy efficiency $\mathcal{E}$ of the network, by adopting the distribution $\Updelta^{\star}$ presented in Table~\ref{table:OptimalDistribution}.  Fig.~\ref{fig:MaximumEfficiency} shows the proposed method optimized for $\mathcal{E}$ as a function of the number of end-devices. As observed, the proposed strategy significantly outperforms the standard DRs, allowing for a more judicious use of energy resources. The goal of this optimization is to achieve a balance between energy consumption — associated with the average ToA, which is a function of the total number of header replicas and payload fragments — and the optimal number of packets successfully received by the gateway. Additionally, the energy efficiency obtained with the distribution presented in Table~\ref{table:OptimalDistributionGoodput}, which maximizes goodput, is illustrated. In this case, standard DR9 outperforms the proposed strategy in terms of energy efficiency. However, it is important to note that DR9 is not superior to the proposed scheme when optimized for goodput (Fig.~\ref{fig:Goodput_devices}). Thus, one can see that a trade-off between goodput and energy efficiency is established by comparing the results from Fig.~\ref{fig:Goodput_devices} with those of Fig.~\ref{fig:MaximumEfficiency}.

\begin{table}[!t]
    \renewcommand{\arraystretch}{1.5}
    \caption{$\Updelta^{\star}$ to maximize $\mathcal{E}$, for $l=10$ bytes [\%].}
    \centering
    \begin{tabular}{c|cccccc}
        \toprule
        $\mathbf{M}$ & $\delta_{1}$ & $\delta_{2}$ & $\delta_{3}$ & $\delta_{4}$ & $\delta_{5}$ & $\delta_{6}$ \\  
        \midrule
        $20000$ & $100$\% & - & - & -  & - & -  \\
        $40000$ & - & $100$\% & - & -  & - & - \\
        $60000$ & - & $100$\% & - & -  & - & - \\
        $80000$ & - & $100$\% & - & -  & - & - \\
        $100000$ & - & $100$\% & - & -  & - & - \\
        $120000$ & - & $100$\% & - & -  & - & - \\
        $140000$ & $15$\% & $85$\% & - & -  & - & - \\
        $160000$ & $80$\% & - & - & -  & - & $20$\% \\
        $180000$ & $85$\% & - & - & -  & -  & $15$\% \\
        $200000$ & $85$\% & - & - & - & - & $15$\% \\
        \bottomrule
    \end{tabular}
    \label{table:OptimalDistribution}
\end{table}

In particular, observing Table~\ref{table:OptimalDistribution}, the optimization that maximizes $\mathcal{E}$ favors configurations with fewer header replicas and lower code rates, which diverges from the strategy to maximize goodput detailed in Table~\ref{table:OptimalDistributionGoodput}.
This is evidenced by the prevalence of configurations $\mathcal{S}_1$ and $\mathcal{S}_2$ up to approximately $M=140000$ end-devices. With a lower number of end-devices, there are fewer collisions, leading to an optimal distribution that employs setups with fewer header replicas and payload fragments to maximize the energy efficiency of the network, although sacrificing goodput. However, as the number of devices increases, the proposed strategy efficiently adjusts the setup mix, adopting a more resilient approach to increase the success of packet delivery by incorporating configuration $\mathcal{S}_6$ into the system. The inclusion of $\mathcal{S}_6$ can be explained by the higher number of collisions and the need to improve the packet success probability while maintaining the overall average energy consumption of the network at limited values. Moreover, note again that configuration $\mathcal{S}_3$ (equivalent to DR9), has never been used in Table~\ref{table:OptimalDistribution}.

\subsection{Influence of the Payload Length}

\begin{figure}[!t] 
\centering
\includegraphics[width=8.8cm]{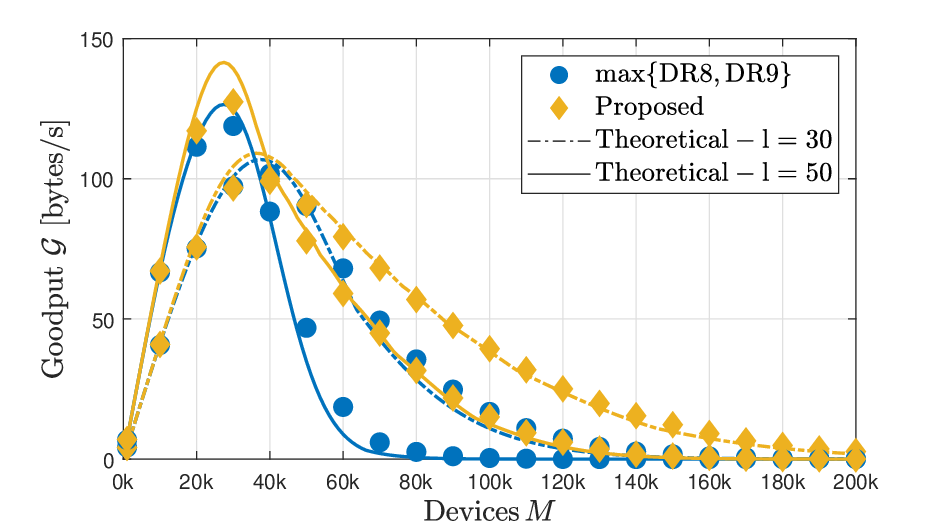}
\caption{Goodput as a function of the number of devices for the proposed method for different payload sizes $l \in \{30, 50\}$ bytes. The maximum performance between standardized DR8 and DR9 is also shown.}
\label{fig:Goodput_devices_payload}
\end{figure}

\begin{figure}[!t] 
\centering
\includegraphics[width=8.8cm]{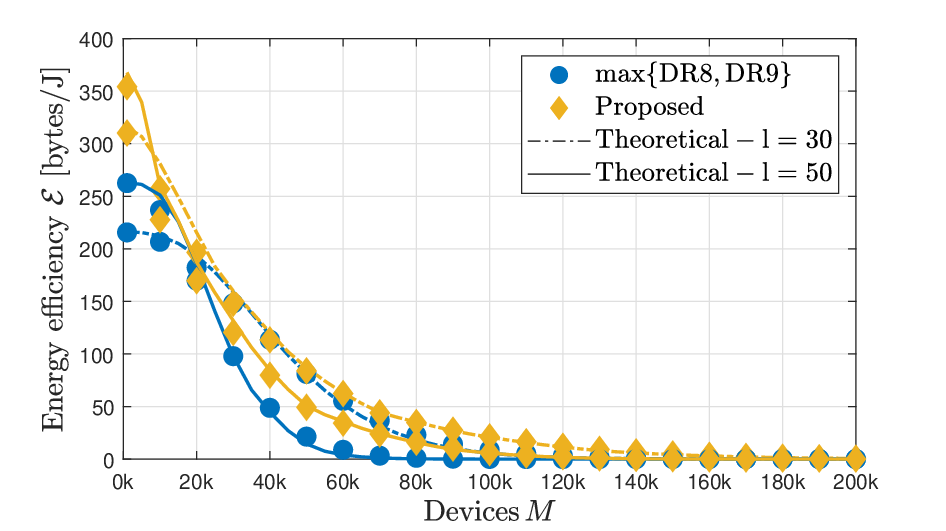}
\caption{Energy efficiency versus the number of devices for the proposed method for different payload sizes $l \in \{30, 50\}$ bytes. The maximum performance between standardized DR8 and DR9 is also shown.}
\label{fig:Efficiency_devices_payload}
\end{figure}

The influence of the payload length on the goodput and the energy efficiency of the proposed scheme are illustrated, respectively, in Fig.~\ref{fig:Goodput_devices_payload} and Fig.~\ref{fig:Efficiency_devices_payload}, for payload sizes $l \in \{30, 50\}$~bytes. 
 Note that, for the sake of conciseness, we compare the proposed method with the maximum between the performances achieved by DR8 and DR9. Even considering such optimistic measure for the standardized DRs, the proposed method presents considerable advantage in terms of goodput and energy efficiency, for both payload sizes.  

Table~\ref{table:OptimalDistributionGoodput_Payload} illustrates the optimal distribution $\Updelta$ to maximize $\mathcal{G}$ for the aforementioned payload values ($\delta_2$, $\delta_3$ and $\delta_4$ are not used, thus omitted), while the optimal distribution from the energy efficiency perspective is shown in Table~\ref{table:OptimalDistributionEfficiency_Payload}. Notably, the setup $\mathcal{S}_3$ is once more not utilized to maximize goodput, being finally used to maximize $\mathcal{E}$ in Table~\ref{table:OptimalDistributionEfficiency_Payload}, but only for a relatively low number of devices.

\begin{table}[!t]
\renewcommand{\arraystretch}{1.5}
\caption{$\Updelta^{\star}$ to maximize $\mathcal{G}$, for $l \in \{30, 50\}$ bytes [\%].}
\centering
\begin{tabular}{c|ccc|ccc}
\toprule
\multirow{2}{*}{$\mathbf{M}$} & \multicolumn{3}{c|}{$l=30$ bytes} & \multicolumn{3}{c}{$l=50$ bytes} \\
 & $\delta_1$ & $\delta_5$ & $\delta_6$ & 
$\delta_1$ & $\delta_5$ & $\delta_6$ \\  
\midrule
$20000$ & - & $100$\% & - & - & $100$\% &  - \\
$40000$ & - & $100$\% & - &  $20$\% & $80$\% & - \\
$60000$ & $35$\% & - & $65$\% & $65$\% & - & $35$\% \\
$80000$ & $60$\% & - & $40$\% & $80$\% & - & $20$\%  \\
$100000$ & $70$\% & - & $30$\% & $85$\% & - & $15$\% \\
$120000$ & $80$\% & - & $20$\% & $90$\% & - & $10$\% \\
$140000$ & $85$\% & - & $15$\% & $95$\% & - & $5$\% \\
$160000$ & $85$\% & - & $15$\% & $95$\% & - & $5$\% \\
$180000$ & $90$\%  & - & $10$\% & $95$\% & - & $5$\%\\
$200000$ & $90$\% & - & $10$\% & $95$\% & - & $5$\% \\
\bottomrule
\end{tabular}
\label{table:OptimalDistributionGoodput_Payload}
\end{table}

\begin{table*}[!t]
\renewcommand{\arraystretch}{1.5}
\caption{$\Updelta^{\star}$ to maximize $\mathcal{E}$, for $l \in \{30, 50\}$ bytes [\%].}
\centering
\begin{tabular}{c|cccccc|cccccc}
\toprule
\multirow{2}{*}{$\mathbf{M}$} & \multicolumn{6}{c|}{$l=30$ bytes} & \multicolumn{6}{c}{$l=50$ bytes} \\
 & $\delta_1$ & $\delta_2$ & $\delta_3$ & $\delta_4$ & $\delta_5$ & $\delta_6$ & 
$\delta_1$ & $\delta_2$ & $\delta_3$ & $\delta_4$ & $\delta_5$ & $\delta_6$ \\  
\midrule
$20000$ & $100$\% & - & - & -  & - & - & - & -  & $100$\% & -  & - & - \\
$40000$ & - & $50$\% & $50$\% & - & - & - & $40$\% & -  & - & -  & $60$\% & - \\
$60000$ & $45$\% & - & - & $55$\% & - & - & $70$\% & -  & - & -  & -  & $30$\% \\
$80000$ & $70$\% & - & - & - & - & $30$\% & $80$\% & -  & - & -  & - & $20$\% \\
$100000$ & $80$\% & - & - & - & - & $20$\% & $90$\% & -  & - & -  & - & $10$\% \\
$120000$ & $85$\% & - & - & - & - & $15$\%& $90$\% & -  & - & -  & - & $10$\% \\
$140000$ & $85$\% & - & - & - & - & $15$\% & $95$\% & -  & - & -  & - & $5$\% \\
$160000$ & $90$\% & - & - & - & - & $10$\% & $95$\% & -  & - & -  & - & $5$\% \\
$180000$ & $90$\% & - & - & - & - & $10$\% & $95$\% & -  & - & -  & - & $5$\% \\
$200000$ & $90$\% & - & - & - & - & $10$\% & $95$\% & -  & - & -  & - & $5$\% \\
\bottomrule
\end{tabular}
\label{table:OptimalDistributionEfficiency_Payload}
\end{table*}

\subsection{Quantization of $\Updelta$}

Due to the prevalence of setups $\mathcal{S}_1$ and $\mathcal{S}_6$ in the optimal distribution $\Updelta$ to maximize $\mathcal{G}$ or $\mathcal{E}$ in the previous analyzes, and with a real deployment in mind, we next consider a simplified scenario where only $\tilde{\Updelta} = \{\delta_1, \delta_6\}$ are available. Furthermore, as the network server must inform the appropriate distribution for joining devices, or for those already connected, ${\tilde{\Updelta}}$ must fit into a typical downlink packet. To achieve this, first note that ${\tilde{\Updelta} = \{\alpha, (1-\alpha)\}}$, where $\alpha$ denotes the probability associated with setup $\mathcal{S}_1$. Hence, it is sufficient for the downlink message from the network server to inform $\alpha$. We will next investigate the impact of quantizing $\alpha$ with a resolution of $b$ bits.

\begin{figure}[ht] 
\centering
\includegraphics[width=8.8cm]{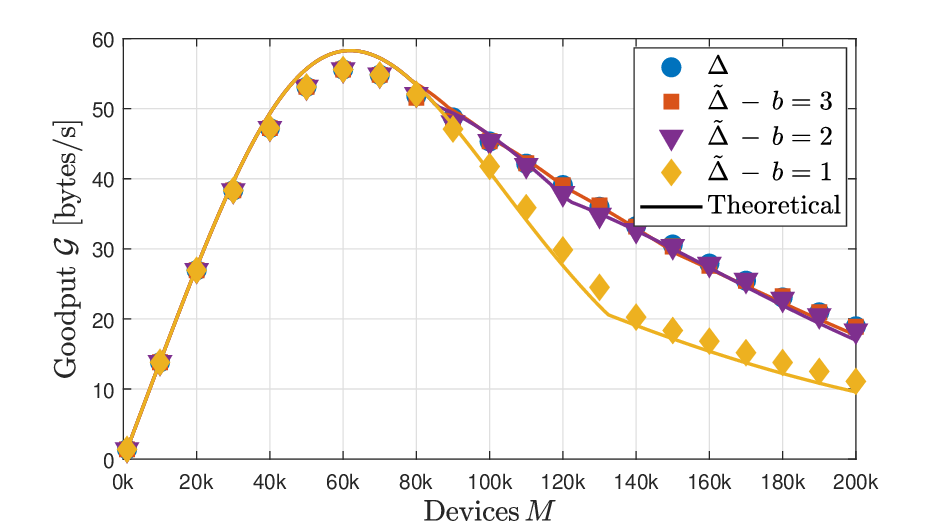}
\caption{Goodput for different quantization resolutions $b$ as a function of the number of devices, for $\Updelta$ and $\tilde{\Updelta}$ optimized for the maximum goodput.}
\label{fig:Goodput_devices_quantization}
\end{figure}

\begin{figure}[ht] 
\centering
\includegraphics[width=8.8cm]{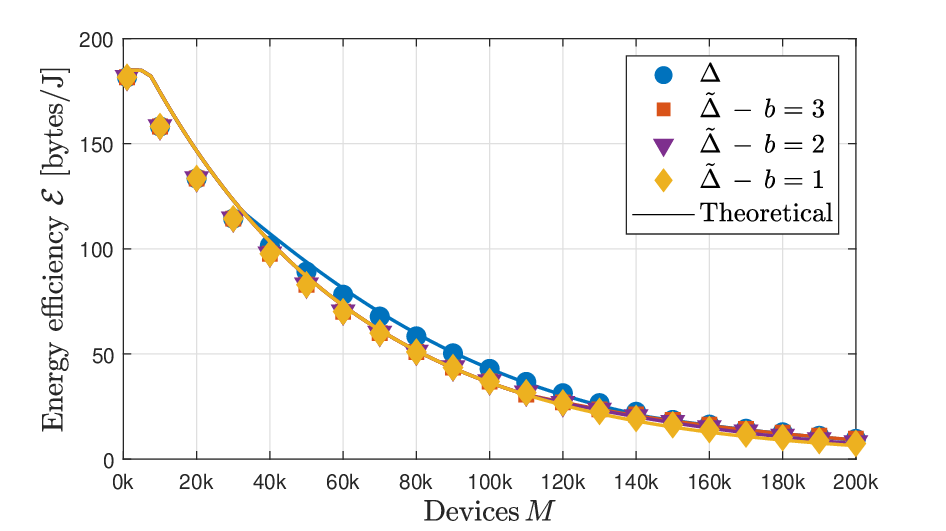}
\caption{Energy efficiency for different quantization resolutions $b$ as a function of the number of devices, for $\Updelta$ and $\tilde{\Updelta}$ optimized for the maximum energy efficiency.}
\label{fig:Efficiency_devices_quantization}
\end{figure}

Figs.~\ref{fig:Goodput_devices_quantization} and~\ref{fig:Efficiency_devices_quantization} illustrate $\mathcal{G}$ and $\mathcal{E}$ for different numbers of bits $b$ as the number of end-devices increases. Notably, a quantization resolution of merely $b=3$~bits is sufficient to achieve near-optimal goodput and energy efficiency. These results demonstrate the feasibility of the proposed approach, as only a minimal number of downlink bits are required from the network server to the end-devices to convey an adequate probability distribution of the setups.

\section{Conclusions}
\label{sec:Conclusions}

We investigated the performance of the LR-FHSS technique using a device-level probabilistic strategy for code rate and header replica allocation, comparing it to the standard DRs. Our results demonstrated that by properly tuning the distribution of header replicas and payload code rates, considerably superior goodput or energy efficiency can be achieved. Notably, data rate DR9, specified in the LoRaWAN standard, was rarely included in our results, while data rate DR8 played a crucial role in the goodput and energy efficiency optimizations. Additionally, regarding downlink transmission from the gateway to end-devices, we found that near-optimal results in terms of goodput and energy efficiency can be attained by transmitting merely 3 bits to inform the appropriate probability distribution, ensuring an adequate probability distribution of the setups with minimal additional downlink traffic load.

\bibliographystyle{IEEEtran}
\bibliography{IEEEabrv,myref}

\end{document}